\documentclass[preprint]{revtex4}

\usepackage{graphicx}
\usepackage{epsfig}
\usepackage{dcolumn}
\usepackage{bm}
\usepackage{amssymb}

\begin{document}

\title{Nonlinear dynamical models from time series}

\author{Jos\'e-Maria Fullana}
\address{Institut Jean Le Rond d'Alembert \\ Universit\'e Pierre et
   Marie Curie\\ e-mail: jose.fullana@upmc.fr}


\begin{abstract}

  We present an optimization process to estimate parameters in
  systems of ordinary differential equations from chaotic time series.
  The optimization technique is based on a variational approach, and numerical studies
  on noisy time series demonstrate that it is very robust and
  appropriate to reduce the complexity of the model. The proposed process
  also allows to discard the parameters with scanty influence on the
  dynamic.

\end{abstract}


\maketitle



\section{Introduction}

The question if given a time series measurements we can identify an
underlying dynamical model and predict its future values begin
certainly with Yule \cite{yule29} and is posed today in several
disciplines as well as economy, geophysics or fluids dynamics.  In
economy research, the short time prediction plays an important role in
financial risk decision; notwithstanding dynamical models for the
different observable are not known. Then the huge quantities of data
we dispose do possible statistical approaches.  Antagonist examples
are seismic inversion, and oil and water research where the model is
in general 
well known, i.e. the wave equation or Darcy like models, but
experimental data are only accessible at the frontier of the studied
region. As a consequence geophysics research has developed powerful
tools of collecting the bulk information. An other interesting and
different example is fluid dynamics because it has a good model, the
Navier Stokes equation, and the possibility of taking data everywhere.
Unfortunately the harvests of the initial conditions are difficult
since requires the measurements of these functions over a
three-dimensional domain.  Then typical experiences in hydrodynamics
produce time series and so, the most
practical situations deal with time series. All these examples are
different outlooks of the same universal question: How can we obtain
dynamical systems from measurements?  It is awful question because we
have an infinite number of models M belonging to a specific class
of functions (radial functions, polynomials, etc.)  which must be
specified in view of the a priori knowledge about the problem, and even when the
model is known it will be parameterized by a set of unknown numbers
$\alpha$.

In this paper we consider a classical problem in nonlinear dynamical
systems: given a noisy time series, we want to capture the underlying
dynamics and, to do that we suppose that it can be modeled by a
coupled system of ordinary differential equations (ODE) parameterized
by a set of numbers $\alpha$.  We propose a constraint minimization to
reduce the model complexity, that is, to find out parameters with
scanty influence on the dynamic ($\alpha_j \sim 0$), and in addition to
reduce the overfitting risk. The order of the model is given by the
number of non zero components of the vector $\alpha$. We apply a
variational approach to compute the derivatives of $\alpha_j$ for a
defined measure ${\cal F}$ and a step descent method to find the
optimal set of parameters.  We will show on chaotic time series that this
technique is robust and able to reconstruct orbits from noisy data.

\section{Identification Method}
\label{sec:ident-meth}

The baseline time series are generated by the following model of $m$ coupled ODE :
\begin{equation}
  \label{model1}
 \frac{dA_i(t)}{dt}-f_i(\alpha,A_1(t),\cdots,A_m(t))=0 \ \ \ \ \ \ \
i=1,\cdots,m
\end{equation}
where the parameter vector $\alpha \in R^n$.  The integration method
is an Euler schema with time step $\delta t=0.01$ to assure the
stability for long time integrations. A Gaussian noise with zero average and standard
deviation $\tau$, ${\cal N} (0,\tau)$,  is then added to the noiseless
signal $A(t)$ to build the ``observed''
data $D(t)$.  Different noisy time series are produced by modifying the
standard deviation $\tau$.  The amount of noise over the
signal is quantified by the signal-to-noise ratio (SNR) 
\begin{eqnarray*}
  SNR = \left( { S_{data} \over S_{noise} } \right)^2
\end{eqnarray*}
where $S$ stands by the root mean square, in particular in our case $S_{noise}=
\tau$. The logarithm relation $10 \log_{10} SNR$ gives the ratio in
$dB$. The ``observed'' data used for computations are
\begin{eqnarray}
  \label{eq:noise}
  D_i (t) = A_i (t) + {\cal N} (0,\tau)  \ \
i=1,\cdots,m
\end{eqnarray}
where we assumed the measurements performed at fixed sampling
time $T < \delta t$. 

To asses the quality of the reconstruction we define a functional
${\cal F} = \sum_1^m {\cal F}_i$ which consists on the addition of the Euclidean
distances between the observed $D_i$ and the reconstructed data $M_i$
on $w$ time windows over a time integration $L$ :
\begin{eqnarray*}
{\cal F}_i= \sum_{w}\int^L_0 || \ D_i(t) - M_i(t) \
||^2 \  \delta(T) \ dt.
\end{eqnarray*}

The reconstructed data $M_i$ are generated by a model $
  M (\alpha)$ at fixed parameters
$\alpha$, and we note that for a free noise series, when $M(t)$ and $D(t)$
coincide both
with $A(t)$ we have ${\cal F} = 0$.
According to the fact that the measurement are performed in a discrete
way we define a delta function $\delta(T)$ related to
the sampling time $T$; for instance $\delta(T)=1$ for multiple of the
sampling time of observed data and zero elsewhere.

We are in presence of an inverse problem, that is to seek for an
optimal set of parameters of a model with respect to a measure ${\cal
  F}(\alpha)$.  The classical approach for a $n$ dimensional problem
is the unconstrained optimization: minimize ${\cal F}(\alpha)$ with
$\alpha \in R^n$. This becomes most of the time a classical least squares approach or
one of its several variations. For a linear model in the parameters
$\alpha$, the cost function is quadratic and there is only one global
minimum. We can estimate directly the derivatives of the model $M$
from observed data $D_i(t)$ but noise prevent an appropriate
evaluation.  We recall that inverse problems are generally
ill-conditioned which is reflected in the lack of robustness face to
noise showed in numerical
simulations~\cite{fullana97,timmer2000parametric}.

This work presents a constrained optimization and we solve it using a
variational approach.  The formal definition of constrained
optimization is the following: minimize ${\cal F}(\alpha)$ with
$\alpha \in R^n$ subject to a constraints $g(\alpha)=0$.  We define
specifically the $g(\alpha)$ as the proposed model $M$ for the
``observed'' data.  We therefore write $g(\alpha)=0$ as
\begin{equation}
  \label{direct}
 \frac{dM_i}{dt}-f_i(\alpha,M_1,\cdots,M_m)=0 \ \ \ \ \ \ \
i=1,\cdots,m
\end{equation}
where $m$ is the dimension of the data series and the fonction
$f_i(\alpha,M_1,\cdots,M_m)$ belonging to some specific class of
functions. Explicit dependency in time is removed for sake of
readability. On each windows $w$ the model is integrated between $0$
and $L$ and the initial conditions are the ``observed'' values at the
beginning, $M_i(0)= D_i(0)$. Within this situation we are close to an initial value
problem in the framework of the multiple shooting approach.

We define the following Lagrangian function
\begin{equation}
  \label{lagrangian}
  {\cal L}_i = {\cal F}_i + \sum_w \int^L \left( 
dM_i/dt-f_i(\alpha,M) 
 \right) V_i \ dt
\end{equation}
where $V=V (\alpha,M,V)$ is the dual variable corresponding to the
constraint or Lagrange multiplier. As the constraint is always verified we
have ${\cal L}_i= {\cal F}_i $
for any choice of $V_i$.  The total variation is then
\begin{eqnarray}
  \label{eq:variation}
  \delta {\cal  L}_i = \delta_\alpha {\cal L}_i  \delta \alpha +
  \delta_V {\cal L}_i  \delta V_i + \delta_M {\cal L}_i  \delta M.
\end{eqnarray}
We observe that the term $\delta_V {\cal L}_i =0$ is equivalent to the
imposed constraints and therefore zero. Provided that $\delta_V {\cal
  L}_i =0$ it is clear that we can computed the gradient explicitly as
a matter of fact  ${\cal L}_i= {\cal F}_i$ implies 
\begin{eqnarray}
  \label{eq:3}
  {  \delta {\cal  L}_i  \over  \delta \alpha   } =  {  \delta {\cal  F}_i  \over  \delta \alpha   } .
\end{eqnarray}
Imposing $\delta_M {\cal L}_i =0$ results in a system for $V$ that
must be integrated backward in time over the length window $L$. For
each window $w$ this leads to the following 
expression 
\begin{equation}
  - \frac{dV_i}{dt}=\delta_M f(\alpha,M,V)+ 2 e_i(t) \delta(T)
\end{equation}
with the boundary condition set at the final time $L$, $V_i(L)=0$
and where $ e_i(t)=D_i(t) - M_i(t)$ is the local error.
 
The gradient of
the components $i$ of the cost function, ${\cal F}_i$, can be write explicitly as
\begin{equation}
   \delta_{\alpha} {\cal F}_i = - \sum_w 
\int^L \delta_{\alpha} f_i
\cdot \  V_i \ 
        dt. \label{grad}
\end{equation}                                    
and finally the gradient for a given parameter $\alpha_j$ is 
\begin{eqnarray*}
   \delta_{\alpha_j} {\cal F} = \sum_i \delta_{\alpha_j} {\cal F}_i 
\end{eqnarray*}

Once the gradient established we perform a descent in the
direction of the gradient of ${\cal F}$.  We apply a {\em
quasi-Newton} method which uses the function gradient at each
iteration \cite{gill81}.  

The optimisation algorithm find an optimal set which
depends on the noise level $\tau$ and on the window length $L$,
$\bar{\alpha} =\bar{\alpha}(\tau,L)$. The
model is evaluated on each window for fixed parameters $\alpha$ for
different $L$ starting from $L=1$.  The $w$ windows of temporal length
$L$ are thus equivalent to the fitting intervals of the multiple
shooting method but the difference being that we do not impose the
matching between solutions at the interval frontier ($t=T$). This exemple has
been well examined in reference \cite{baake1992fitting} in two cases :
when the model is available and when we specify only its class.

\begin{figure}[htb]
  \centering
  \includegraphics[width=0.5\textwidth]{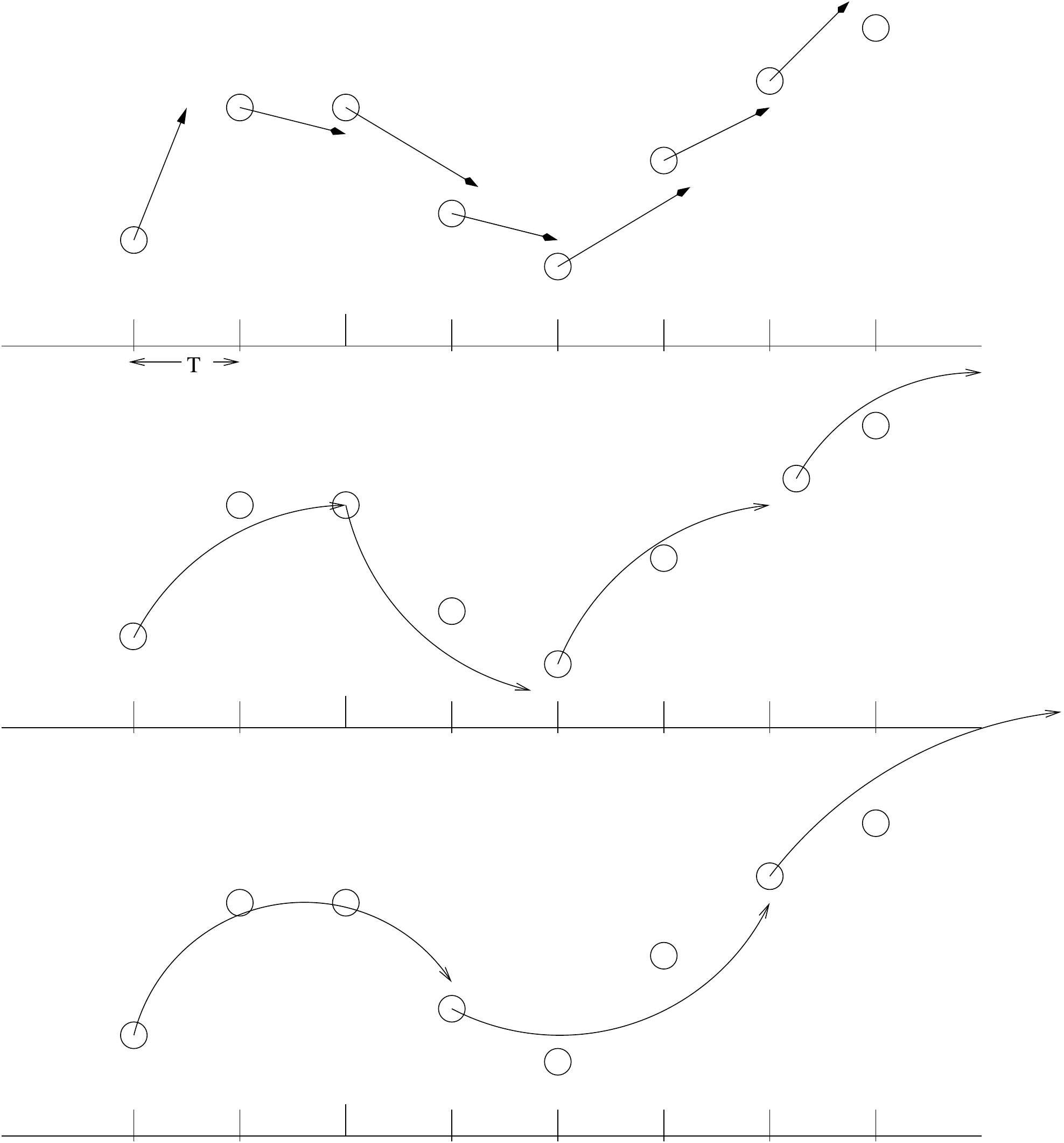}
  \caption{Scheme for the optimisation process for $L=1, 2, 3$ (from
    the top to the bottom). }
  \label{fig:0}
\end{figure}

The Figure~\ref{fig:0} shows a schematic illustration of the process
for $L=1,2,3$. We paid attention to a specific data (the third one
beginning from the left), we observe that it is used on the window :
(i) for $L=1$ as initial condition, (ii) for $L=2$ as the final
condition, and (iii) for $L=3$ as internal data. Then, it is
straightforward of concluding that data are used more that once in the
process in contrast to the multiple shooting method, and that ``window
overlap'' enhances the statistics. To improve yet the statistic we
repeat the procedure over a large number of probes $np$ of length $w
L$. Using the optimal values from each probe we compute the average
value ${\bar \alpha}$ and the standard deviation $\sigma =
\sqrt{\sum_{k=1}^{np} ({\bar \alpha} - \alpha^{opt}_{k} )^2}$.  We can
apply this procedure to experimental data by splitting experimental
data series in probes of size $wL$.

The Figure \ref{fig:1} shows the variation of three parameters ${\bar
  \alpha}$ as function of the window length $L$ for the Lorenz
equation with noise equal to 14.53 $dB$ $(\tau = 1.5)$. In this
case we have $np=10^3$ and the probe size $wL$ is set to $10^3$. We
will discuss the Lorenz model and the figure in the next
section. The standard deviation $\sigma$ is represented by horizontal
ticks. Note that the average values converge from $L=25-30$, in term
of optimisation we can infer that no more information is available.

Finally parameters $\alpha$ with small mean values and weak ratio ${
  {\bar \alpha} \over \sigma } $ are discarded and this determines the
end of the first stage of the optimisation process. The ratio ${ {\bar
    \alpha} \over \sigma } $ is called reliability which is an
estimation of the statistical reliability (the difference is that we
known the average value ${\bar \alpha}$ instead of the actual
value). A central point is how do we quantify it. We decide that
$\alpha_{opt} < eps$ are discarded and we set $eps = 0.02$ which is
quite arbitrary. We argue that if the parameters $\alpha$ are of the
order of 1 this $eps$ implies that the discarded parameters are at least
around 2\% of the keeping ones.  On the contrary in a case with
parameters $\alpha$ of the order of less than 1 we have to re-define
this cutoff. Next, for the reliability the criterion is less
arbitrary as long as  parameters with $\log_{10} ({ {\bar \alpha} \over \sigma }
)< 0$ are considered weak, the cutoff is then $ {\bar \alpha} \sim
\sigma $.

We keep parameters which in general, have high reliability, small
$\sigma$, and are different from zero.  We reduce by the procedure the number of
parameters to avoid overfitting.  In the following stages we use the
original model with a smaller number of parameters and the recursive
procedure stops when the number of parameters does not change anymore. In
practice three or four stages are suffisant to finish the optimisation
procedure.

\section{Applications and Discusion}
\label{sec:applications}

The strategy of model construction and parameter selection presented
above is applied on a particular class of functions, a full 2d-order polynomial
taking into account the squares and the cross products for $m=3$ :

\begin{eqnarray*}
  dM_1(t)/dt= & \alpha_1 M_1(t)  + \alpha_2 M_2(t)  +\alpha_3 M_3(t) + \alpha_4 M_1(t) M_2(t) + \alpha_5 M_1(t) M_3(t) + \\
  &       \alpha_6 M_2(t) M_3(t)  + \alpha_7 M_1^2(t)  + \alpha_8 M_2^2(t)  + \alpha_9 M_3^2(t)  
 \end{eqnarray*}
 \begin{eqnarray}
\label{eq:model}
dM_2(t)/dt=& \alpha_{10}M_1(t)  + \alpha_{11}  M_2(t)  + \alpha_{12} M_3(t) + \alpha_{13} M_1(t) M_2(t) + \alpha_{14} M_1(t) M_3(t) + \\
&       \alpha_{15} M_2(t) M_3(t)  + \alpha_{16} M_1^2(t)  + \alpha_{17} M_2^2(t)  + \alpha_{18} M_3^2(t)  
\nonumber
 \end{eqnarray}
\begin{eqnarray*}
   dM_3(t)/dt=& \alpha_{19}M_1(t)  + \alpha_{20}  M_2(t)  + \alpha_{21} M_3(t) + \alpha_{22} M_1(t) M_2(t) + \alpha_{23} M_1(t) M_3(t) + \\
&       \alpha_{24} M_2(t) M_3(t)  + \alpha_{25} M_1^2(t)  + \alpha_{26} M_2^2(t)  + \alpha_{27} M_3^2(t).
 \end{eqnarray*}
 The
 vector of parameters $\alpha$ is composed of 27 values.  The
 integration method is an Euler schema with time step $\delta t=0.01$
 and the sampling time $T=0.1$. For statistics we use $np=10^3$ and
 the probe size $wL$ is set to $10^3$. We use the same model ($M=A$) to generate the
 noisy series. We note that the chaotic Lorenz model \cite{lorenz63} and the
 Roessler model \cite{rossler1976equation} belong to this class.

We apply therefore the optimisation process to both models. 
Chaotic data series from the Lorenz model are built using
equation~(\ref{eq:model}) with the following seven non-zero $\alpha$
components :  $\alpha_{1} = -10.,  \alpha_{2} = 10., \alpha_{10} =
 28. , \alpha_{11} = -1., \alpha_{14} =  -1. , \alpha_{21} =
-2.666, \alpha_{22} = 1. $. In order of obtaining the ``observed'' data
we add some amount of noise (equation~(\ref{eq:noise})) and we pick
data at sampling times. 

The set of optimal parameters will indeed depend on the noise
level $\tau$ and on the window length $L$ and, in addition, the
functional ${\cal F}$ will become more and more non linear as long as
$L$ growths \cite{henning2004nonlinear}. With this in mind we monitor
the $\bar{\alpha}$ as a function of the window length $L$ starting
from $L=1$. The case $L=1$ is particular, because in a strict sense
the numerical evaluation of the derivatives of the model
(\ref{eq:model}), $dM_1(t)/dt$ by the way, are almost the same using either
$\delta t$ or $T$. It easily follows that for $\delta t = T$ and $L=1$
the optimisation problem becomes a classical least square, then the
functional function ${\cal F}$ is quadratic and the solution unique
and moreover we can compute the derivatives explicitly from ${\cal F}$ without
doing a variational approach.

\begin{figure}[htb]
  \centering
  \includegraphics[width=0.5\textwidth,angle=-90]{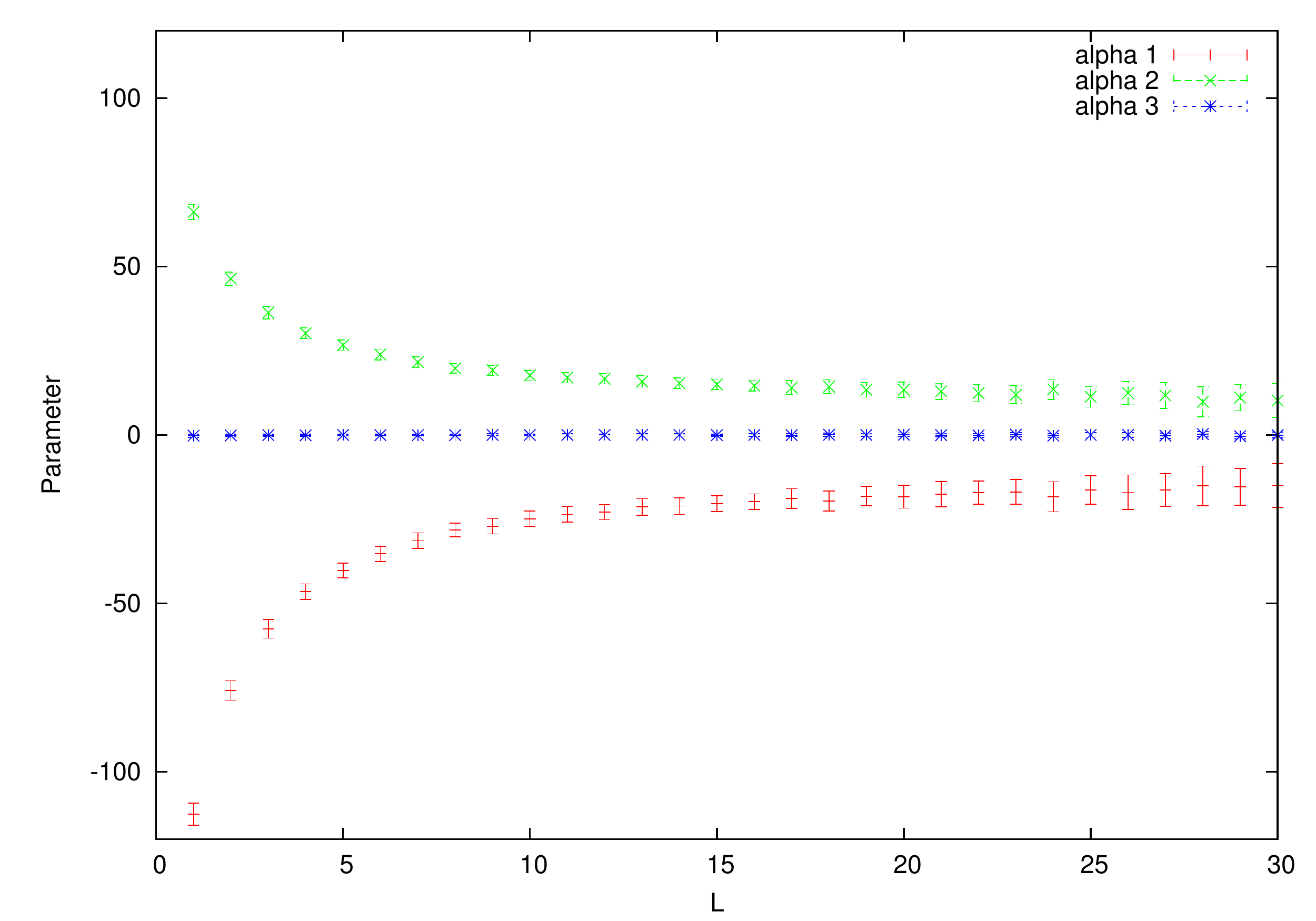}
  \caption{Optimal $\alpha$ as function of the length window $L$ for a
    Lorenz model with noise level around of 14.53  $db$.}
  \label{fig:1}
\end{figure}

The optimal parameters of window length $L=1$ are used as initial guess for the
window length 
$L=2$ and we continue until convergence with the window size $L$. The key
features of the process are showed in Figure~\ref{fig:1} for three parameters $\alpha_1,
\alpha_2, \alpha_3$ where the standard deviation $\sigma$ is
represented by horizontal ticks. 

\begin{figure}[htb]
  \centering
  \includegraphics[width=0.6\textwidth,angle=-90]{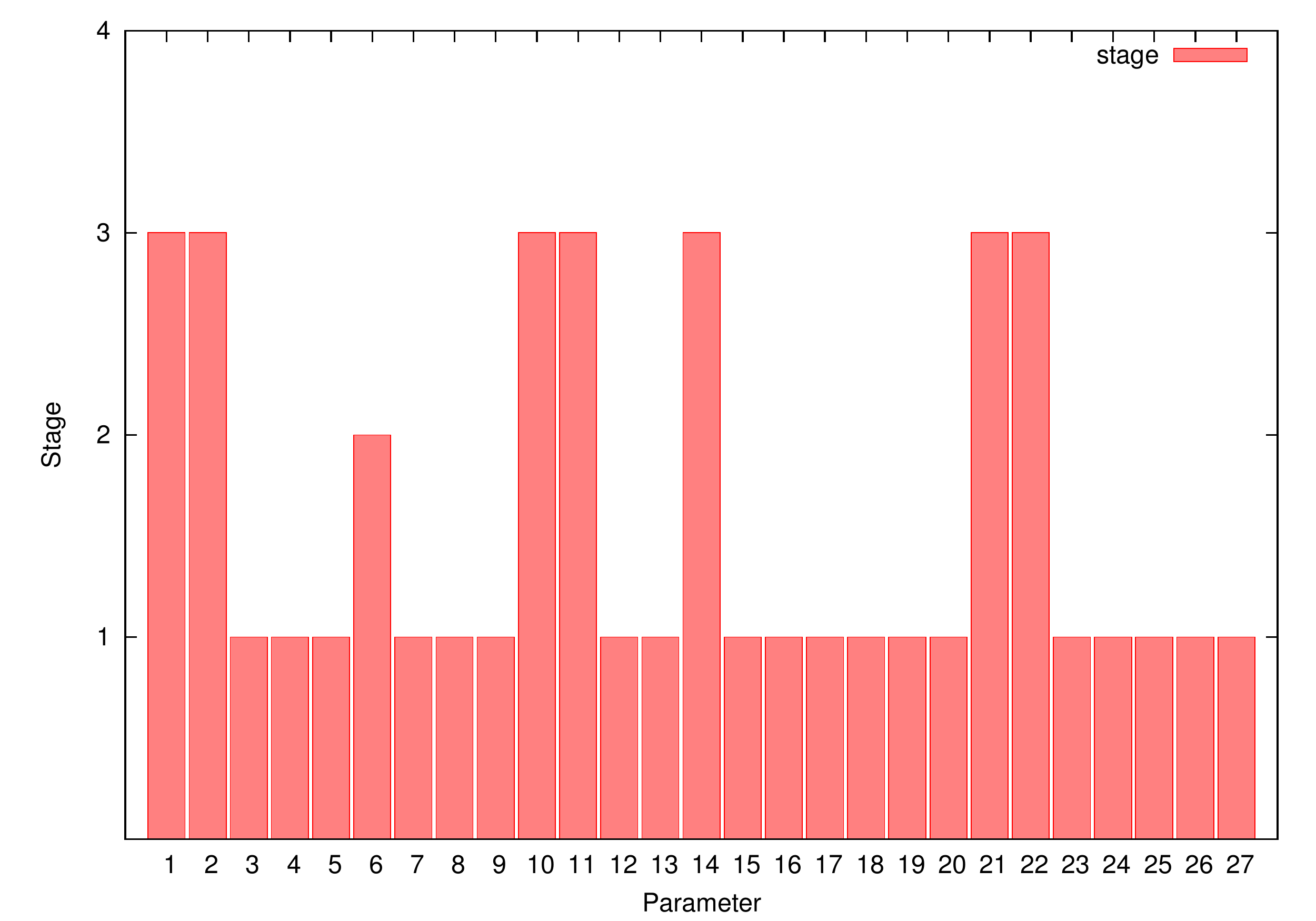}
  \caption{Stages of the optimization process for each parameter
    $\alpha_i$, $i=1 \dots 27$ of the model. Same case than
    Figure~\ref{fig:1} . See text for details. }
  \label{fig:2}
\end{figure}

The level of noise is high, $\tau =1.5$ in
equation~(\ref{eq:noise}), which is around of 14.53 $dB$. Note that
the average values converge from $L=25-30$. Pay attention to the $y$
axis scale, the parameter $\alpha_3$ is not but fluctues around zero, its reliability is
low and it is discarded at the first stage as is shown in
Figure~\ref{fig:2}.

\begin{figure}[htb]
  \centering
  \includegraphics[width=0.5\textwidth,angle=-90]{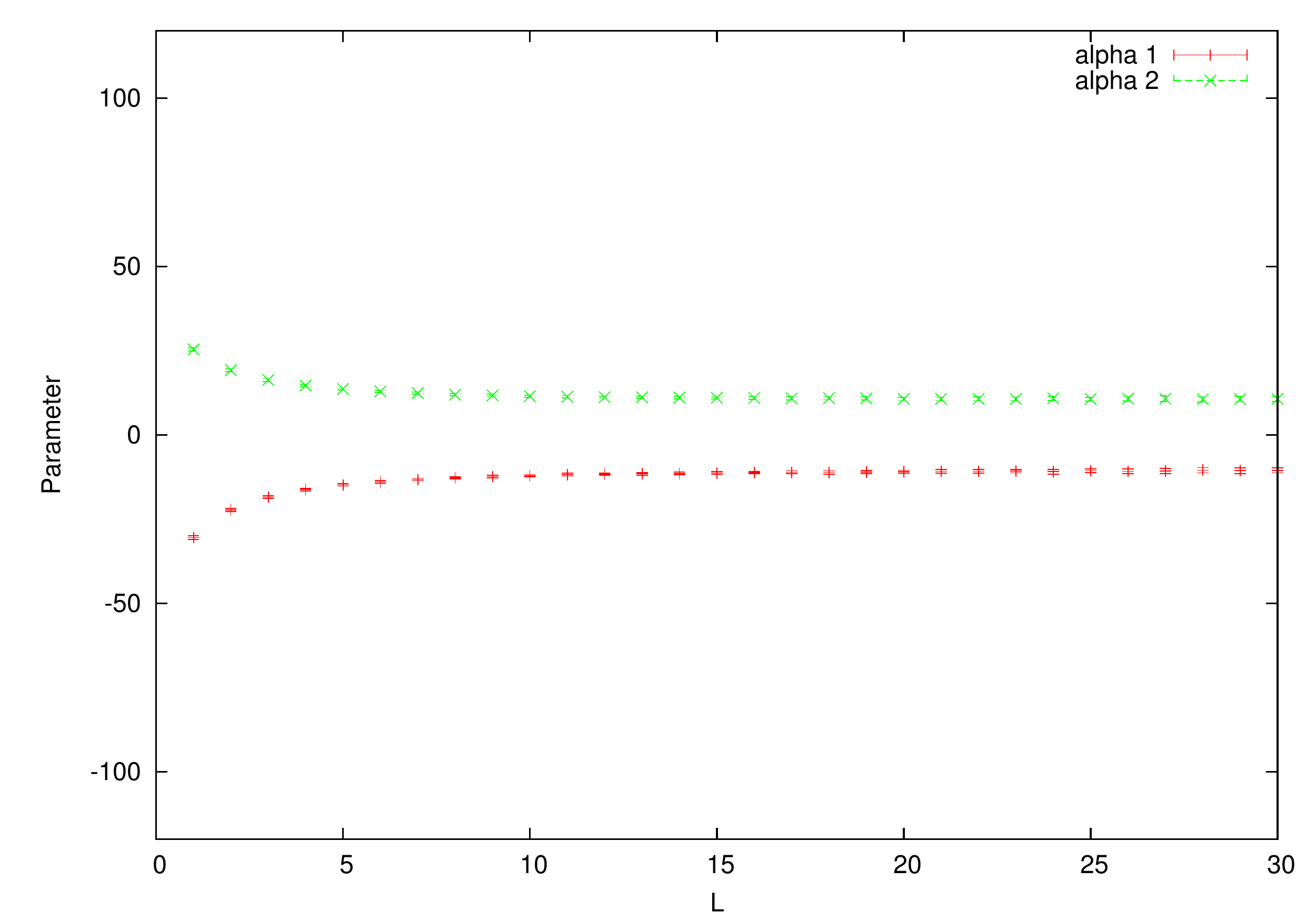}
  \caption{Same than Figure~\ref{fig:1}. Note that $\alpha_3$ has been discarded at the stage 1.}
  \label{fig:3}
\end{figure}

The Figure~\ref{fig:2} presents then the process of discarding
parameters : the $y$ axis say if given parameter is dropped off or
not a that stage. In some detail we can see that parameter 1
($\alpha_1$) is stilll present at the end of the process (stage 3),
and conversely parameter 3 ($\alpha_3$) disappear in the first run
(stage 1). As noted earlier parameters $\alpha$ with small mean values
and weak reliability are discarded.  In the Figure~\ref{fig:3} we show
the equivalent of Figure~\ref{fig:1} but at the end of the process
(stage 3).

We note that the convergence is done before $L=25-30$ and both the
optimal parameters and dispersions are less affected by noise (in
particular compare optimal values for small window length $L$ in
Figure~\ref{fig:1}). 

The Table~\ref{tab:probes} shows optimal values and standard deviation for several
noisy series ($\tau = 0.5, 1.0, 1.5, 2.0, 2.5$).

\tiny
\begin{flushleft}
\begin{table}[htb]
\scalebox{0.8}{
  \begin{tabular}{c||cc||cccc}
$\alpha$ & $\tau = 0.5 (10^2)$   & $\tau = 0.5 (10^3)$  &
$\tau = 1.0 (10^3)$ & $\tau = 1.5 (10^3)$  & $\tau = 2.0 (10^3)$ &
$\tau =2.5 (10^3)$ \\ \hline
1 & -10.359 $\pm$ 0.221  & -10.374 $\pm$ 0.247 & -10.578 $\pm$ 0.314  & -10.664 $\pm$ 0.404  & -10.673 $\pm$ -0.509  & -10.591 $\pm$ 2.801 \\
2 & 9.955 $\pm$ 0.180  & 9.956 $\pm$ 0.202 & 9.995 $\pm$ 0.248  & 10.042 $\pm$ 0.291 & 10.149 $\pm$ 0.378  & 10.084 $\pm$ 3.658 \\
10 & 28.673 $\pm$ 0.572  & 28.706 $\pm$ 0.549 & 29.127 $\pm$ 0.779  & 29.297 $\pm$ 0.984 & 29.332 $\pm$ 1.202  & 29.601 $\pm$ 5.653\\
11 & -0.954 $\pm$ 0.075  & 0.952 $\pm$ 0.078 & -0.968 $\pm$ 0.110  & -0.991 $\pm$ 0.147 & -1.028 $\pm$ 0.312  & -0.168 $\pm$ 4.504\\
14 & -1.021 $\pm$ 0.019  & -1.022 $\pm$ 0.018 & -1.034 $\pm$ 0.027  & -1.041 $\pm$ 0.036& -1.043 $\pm$ 0.051  & -1.042 $\pm$ 0.162\\
21 & -2.608 $\pm$ 0.020  & -2.607 $\pm$ 0.022 & -2.603 $\pm$ 0.033  & -2.598 $\pm$ 0.043 & -2.589 $\pm$ 0.062  & -2.495 $\pm$ 0.474 \\
22 & 1.023 $\pm$ 0.021  & 1.023 $\pm$ 0.023 & 1.036 $\pm$ 0.035  & 1.051 $\pm$ 0.048 & 1.047 $\pm$ 0.078  & 1.167 $\pm$ 1.120 \\
26 & -- & -- & -- & -- & 0.054 $\pm$ 0.085  & 0.094 $\pm$ 0.457 
\end{tabular}
}
\caption{Optimal parameters for the Lorenz model, average values and
  standard deviation.
  (columns two and three) Noise level is $\tau = 0.5$ for $np=10^2$ and
  $np=10^3$. (columns four to seven) Noise leves are  $\tau = 1.0,
  1.5, 2.0, 2.5$ ($18.06, 14.53, 12.04, 10.10$
  in $dB$ using the baseline signal of $A_1$) for $np=10^3$
  probes. }
\label{tab:probes}
\end{table}
\end{flushleft}
\normalsize

Each column shows the average value $\bar{\alpha}$ and its standard
deviation. Columns two and three illustrate their behavior for the
same noise level $\tau = 0.5$ and for two probes number $np=10^2$ and
$np=10^3$, we can see that statistic does not change too much, we conserve for
that reason $np=10^3$ for the subsequent computations. Columns four to
seven present the evolution of the numerical findings for noise levels
$\tau = 1.0, 1.5, 2.0, 2.5$ which is in $dB$ $18.06, 14.53, 12.04,
10.1$ using the noiseless signal $A_1$. At $\tau = 2.0$ the parameter
26 ($\alpha_{26})$ does not disappear and is still present at the end of
the optimization process. We note that the other parameter are not
affected by its presence and the numerical reconstruction (not shown) using the
ODE system~(equation (\ref{eq:model})) does not differs from the
original time series.

\begin{figure}[htb]
  \centering
  \includegraphics[width=0.8\textwidth]{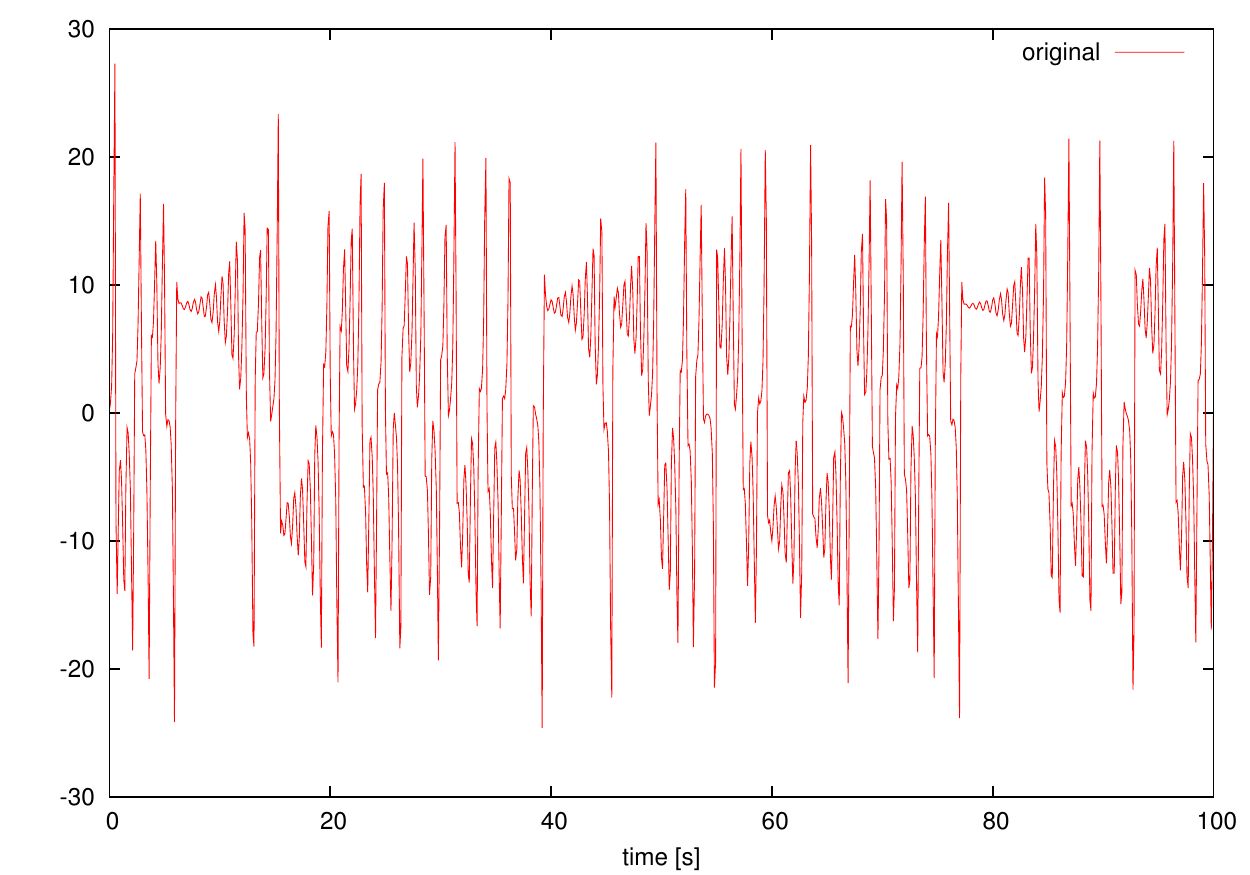}

  \includegraphics[width=0.8\textwidth]{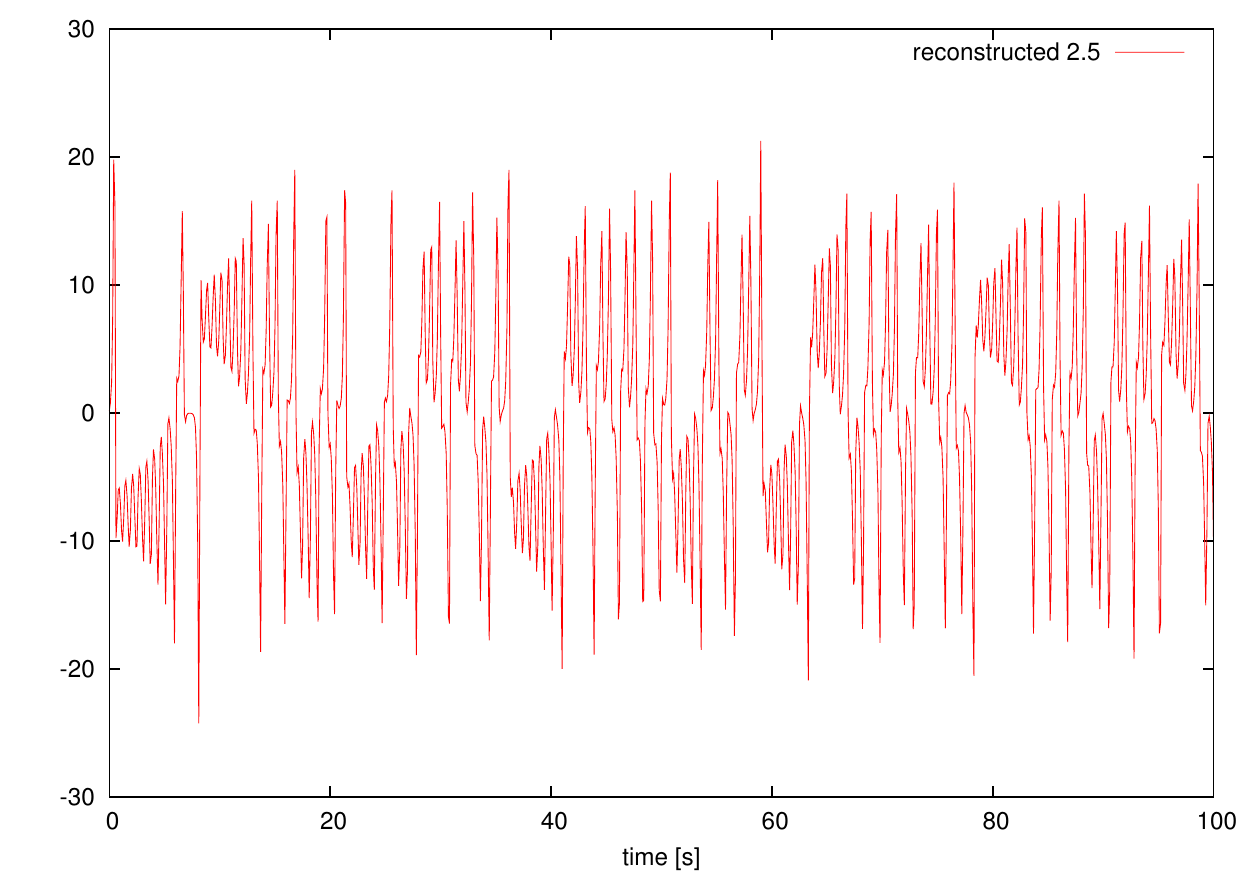}

  \caption{Lorenz model :   (top) original temporal series $A_2(t)$ and
    (bottom ) reconstructed one $M_2(t)$ for $\tau=2.5$.}
  \label{fig:4}
\end{figure}

On the contrary, the next result ($\tau = 2.5$) shows that parameter
11 is seriously affected varying of one order of magnitude, from $\sim
-1$ to $\sim -0.1$. Parameters coming from the linearized system as
$dM_2(t)/dt \sim \alpha_{11} M_2(t)$ among others are really difficult
to obtain. These terms give, in a first approximation exponential
solutions behaving like $M_2(t) \sim e^{\alpha_{11} t}$. When noise is
added we shadow the orbits and a lot of them are equivalent, we are in
the conditions of the ``shadowing'' lemma (\cite{farmer91a} and
references therein) but for an inverse problem : under these
conditions there are not
enough  information to provide accurate estimates of the parameter values
and then the optimization algorithm is not able to separate
contributions coming from linear or nonlinear terms \cite{Smith2010}. Even though the
parameters are affected by noise the reconstruction using
equation~(\ref{eq:model}) shows the typical ``strange'' Lorenz
attractor and we can see in Figure \ref{fig:4} that the time series
for the variable $M_2(t)$ is very similar to the original
$A_2(t)$. Nevertheless a close examination of the ``burst'' regions
show us that the reconstructed one is less sharp. This characteristic
is governed by the coefficient $\alpha_{11} $ which, as mentioned
above, is not well evaluated.

\begin{figure}[htb]
  \centering
  \includegraphics[width=0.8\textwidth]{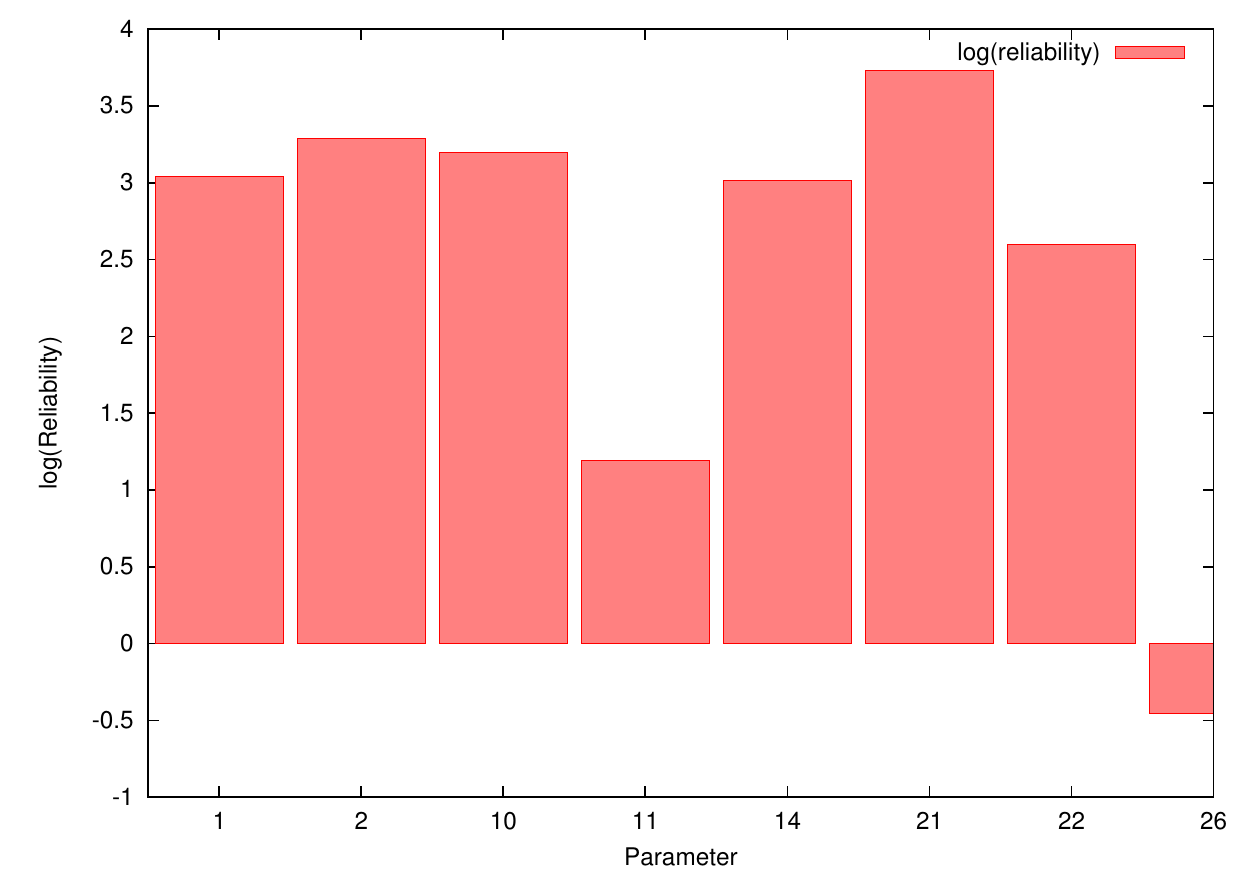}  
\caption{Logarithm of the reliability for coefficients of
  Table~\ref{tab:probes}, column 6.}
  \label{fig:5}
\end{figure}

Coming back to column 5 ($\tau = 2.0$) and examining the parameter 26
($\alpha_{26}$) we remark that even if it is larger than the cutoff
value "eps" its standard deviation is also quite large, then its reliability
is poor. Figure~\ref{fig:5} presents the logarithm of the reliability
of all parameters, this result sugests that satisfying 
 both criteria (small
mean values and weak ratio ${ {\bar \alpha} \over \sigma }$) conjointly is too
restrictive. We restart the process using either small
mean values OR weak ratio ${ {\bar \alpha} \over \sigma }$, the ``OR''
condition for the rest of the paper.

The Table~\ref{tab:2} shows numerical results for $\tau=2.0$ and
$\tau=2.5$ for the ``OR'' condition, in both cases the optimal values are closed to the actual
ones and their resist quite well to the noise, except for the parameter 11
which is becoming not reliable, ${ {\bar \alpha_{11}} \over \sigma_{11} } \sim 1$.

\begin{table}[htb]
  \centering
  \begin{tabular}{c|cc}
$\alpha$  & $\tau = 2.0,
np=10^3$ & $\tau =2.5, np=10^3$ \\ \hline
1 & -10.712 $\pm$ -0.639   & -10.724 $\pm$ -0.868 \\
2 & 10.067 $\pm$ 0.426   & 10.124 $\pm$ 0.747 \\ 
10 & 29.492 $\pm$ 1.629   & 29.699 $\pm$ 2.630 \\ 
11 & -1.021 $\pm$ -0.351   & -1.124 $\pm$ -1.021 \\ 
14 & -1.052 $\pm$ -0.061  & -1.066 $\pm$ -0.098 \\ 
21 & -2.590 $\pm$ -0.125  & -2.566 $\pm$ -0.224 \\
22 & 1.069 $\pm$ 0.074  & 1.105 $\pm$ 0.167 
\end{tabular}
\caption{Optimal coefficients for the Lorenz model using the ``OR''
  condition (see text). Noise values are $\tau=2.0$ and
$\tau=2.5$.}
\label{tab:2}
\end{table}

We apply now the ``OR'' condition to the Rossler model \cite{rossler1976equation} .
Observed data for the Rossler model are generated by
equations~(\ref{eq:model}) with parameter $\alpha= \{ \alpha_2 = -1.0,
\alpha_3 = -1.0, \alpha_{10} = 1.0, \alpha_{11} = 0.38, \alpha_{19} =
0.3, \alpha_{21} = -4.5,\alpha_{23} = 1.0 \}$ plus a noise value as
mentioned above. Noise levels are $\tau = 0.2, 0.4, 0.8, 1.0$ or in
$dB$, $23.72, 17.70, 11.68, 9.74$ using the noiseless signal $A_1$.

\begin{table}[htb]
  \centering
  \begin{tabular}{c|cccc}
$\alpha$  & $\tau = 0.2,
np=10^3$ & $\tau =0.4, np=10^3$  & $\tau =0.8, np=10^3$ & $\tau =1.0, np=10^3$\\ \hline
2 & -1.000 $\pm$ 0.005  & -0.993 $\pm$ 0.012 & -1.002 $\pm$ 0.016   & -1.003 $\pm$ 0.015\\
3 & -0.998 $\pm$ 0.005  & -0.994 $\pm$ 0.010  & -0.983 $\pm$ 0.020   & -0.987 $\pm$ 0.022\\
10 & 0.998 $\pm$ 0.006  & 0.995 $\pm$ 0.013  & 0.979 $\pm$ 0.022 & 0.979 $\pm$ 0.027\\
11 & 0.375 $\pm$ 0.004  & 0.365 $\pm$ 0.008  & 0.328 $\pm$ 0.033   & 0.329 $\pm$ 0.054 \\
19 & 0.313 $\pm$ 0.006  & 0.412 $\pm$ 0.034 & 0.442 $\pm$ 0.087   & 0.433 $\pm$ 0.141\\
21 & -4.476 $\pm$ 0.040 & -4.476 $\pm$ 0.077 & -4.296 $\pm$ 0.168   & -4.329 $\pm$ 0.215  \\
23 & 0.992 $\pm$ 0.010 & 0.974 $\pm$ 0.018 & 0.920 $\pm$ 0.055   & 0.927 $\pm$ 0.081
\end{tabular}
\label{tab:3}
\caption{Optimal coefficients for the Rossler model using the ``OR''
  condition (see text). Noise values are $\tau = 0.2, 0.4, 0.8, 1.0$
  ($23.72, 17.70, 11.68, 9.74$ in $dB$). }
\end{table}

We use the model~(\ref{eq:model}) to identify the unknown parameters and we look
the behavior of the optimal parameters $\alpha$. Table \ref{tab:3}
shows the numerical findings, the first point is that the procedure is
able to determine the important model's parameters even in presence of
high amount of noise. Second, that the standard deviations increase
linearly with the noise level.  Third, that the optimal values resist quite
well to noise excepted parameter 19 $\alpha_{19}$ which has around
$30 \%$ of error. This parameter corresponds to the term $dM_3(t)/dt \sim
\alpha_{13} M_1(t)$ (see the
model (\ref{eq:model})) and is directly responsible for the aperiodic
and rough burst in coordinate 3 on the Rossler model. This burst
dynamics is not easy of capturing because very sharp and short (typical
half time peak is around of 1 sec, which is $L=10$). 

Numerical optimisations show that the optimal values are in excellent
accord and the reconstructed orbits (not shown) are in agreement with
the original ones.

\section{Conclusion}
\label{sec:conclusion}

We have presented an optimization procedure able to retain the
important parameters for a mathematical model given a noisy time
series. We have applied the procedure to two chaotic series from the
Lorenz and the Rossler models. We have demonstrated using an ODE
system that the procedure (i) is appropriate to reduce the complexity
of the model, (ii) is powerful to make a parameters estimation, and
(iii) is very robust against noise. We also observe that we can reduce
the time integration, and in the limit $\delta t \rightarrow 0$ we
could compute the continuos parameters of a system.

In our numerical examples we know the actual number and value of the
model parameters, this is an important help to decide what criterium
we have to use. We have shown that using both criteria, small mean
values and weak ratio ${ {\bar \alpha} \over \sigma }$, conjointly is
too constrain. When at least one criterium is fulfilled we have proved
that the optimization procedure is improved.  In a case of unknown
data series we have to ameliorate the procedure, an interesting way
could be to compute the covariance matrix $ ({ \partial^2 F
  \over \partial^2 {\alpha} })^{-1}$ and analyse the eigenvalues to
display linear combinations.

We can yet improve the procedure working on the initial conditions for each
window : the choice $M_i(0)=D_i(0)$  is may be not the optimal,
numerical tests show that using $M_i(0)=A_i(0)$ (which is in fact
impossible on real data) the optimal results are closest to the actual values. Then
an issue could be either to make a some kind of average process in the
neighbor of the initial conditions 
or add some constraints as well as in the multiple shooting approach.  We could
define the initial conditions $M_i(0)$ as parameters to optimize as
in reference \cite{Fullana97b} but the number of unknowns will become
too important, adding by the way $m \ w$ parameters ($w$ being the window
number) to the optimization problem.  

However many questions are still open : the application to real data
or the tractability for applying this procedure to systems including
unobserved data series (when data dimension is greater that model
dimension).  The present procedure should be effective for different
problems from classical or less classical parameter identification
\cite{timmer2000parametric,PhysRevE.85.036201} to
control/synchronization applications \cite{Yu:2008,Peng:2011}. In
particular an extension to high degree polynomial class is
straightforward and could be effortless applied to neuronal or
electrical circuits \cite{Schumann-Bischoff:2011}.


\end{document}